\documentclass[pra,twocolumn,superscriptaddress,amsmath,amssymb,showpacs,floatfix]{revtex4}
\usepackage{graphicx}
\usepackage{dsfont}
\usepackage{braket}
\usepackage{amsmath}
\usepackage{color}
\usepackage{subcaption}
\usepackage{lipsum}
\usepackage{hyperref}
\usepackage{ragged2e}
\usepackage{placeins}
\usepackage[normalem]{ulem}
\begin{document}
\title{Entanglement Structure Across $\mathbb{Z}_n$ Phase Transitions in 1D Rydberg Atom Arrays}
\author{Hyeonjun Yeo}
\affiliation{Department of Physics and Astronomy, Seoul National University, Seoul 08826, Korea}
\author{Kabgyun Jeong}
\affiliation{Institute of Computer Technology, Seoul National University, Seoul 08826, Korea}
\author{Hyunchul Nha}
\email{hnha@utep.edu}
\affiliation{Department of Physics, University of Texas at El Paso, El Paso, Texas 79968, USA}

\date{\today}
\begin{abstract}
Multipartite quantum entanglement plays a crucial role in the emergence of different quantum phases and their transitions in quantum many-body systems. It is of general interest to know what sort of analysis on quantum entanglement can bring us a profound insight to understand the rich dynamics of quantum many-body systems. In this work we study the characteristics of quantum entanglement in relation to $\mathbb{Z}_n$-ordered phases emerging under a varied strength of 1-dim Rydberg interaction. We propose an approach based on the structure of pair-wise entanglement across the Rydberg chain using two-qubit concurrence as an entanglement measure. We define an entanglement-structure factor via Fourier analysis of total concurrence at each site and address $\mathbb{Z}_n$ phase transitions in comparison with the conventional order-parameter based on local density, i.e. magnetization.  We also discuss how the required two-qubit concurrence can be measured in analog Rydberg atom arrays using site-selective erasure and parametrized laser pulses. Our investigation suggests that an entanglement-structure-based approach can provide a powerful tool in analyzing symmetry-breaking in quantum phase transitions. 

\end{abstract}

\maketitle

\section{Introduction}

Quantum phase transition is a crucial aspect of quantum many-body physics that manifests collective behavior among constituent particles or subsystems \cite{sondhi1997continuous,sachdev1999quantum,sachdev2011quantum}. Unlike classical phase transitions driven by thermal fluctuations, quantum phase transitions are driven by quantum fluctuations at zero temperature in conjunction with the change of external parameters controlling the Hamiltonian of the system. Similar to classical phase transitions, many quantum phase transitions have been classified by their scaling behavior of order parameters that manifests spontaneous symmetry breaking. There are also special phase transitions that cannot be explained by symmetry breaking such as topological phase transitions \cite{kitaev2006topological,wen2017colloquium}.

Quantum entanglement, a fundamental non-local property of quantum systems, has been extensively studied for its role in revealing quantum features. For instance, a quantitative measure of entanglement based on von Neumann entropy has been used to detect quantum phase transitions \cite{osborne2002entanglement1}. In Refs. \cite{osborne2002entanglement2} and \cite{osterloh2002scaling}, two-qubit concurrence was adopted to exhibit the scaling behavior of its derivative at a critical point in translationally symmetric Ising or Heisenberg XY chains, as expected from conformal field theory involving symmetry breaking. In addition, scaling laws of entanglement entropy have been intensively studied across quantum phase transitions \cite{eisert2010colloquium, bianchi2022volume}.
However, entanglement entropy does not deliver the whole picture of quantum correlations among different parties in quantum many-body systems. In a sense, it measures the entanglement between two divided subsystems without addressing multipartite structure explicitly. Furthermore, it is not a directly measurable quantity in experiment. Therefore, while the analysis based on entanglement entropy offers valuable insight, it is important to develop a tool that can probe the role of quantum many-body entanglement in emerging quantum dynamics desirably in an experimentally accessible form. 

In this paper, we propose an approach analyzing the entanglement structure of a quantum many-body system that can be useful to characterize quantum phase transitions. 
Some previous works also addressed the relationship between two-qubit concurrence and symmetry breaking but in a different context  \cite{syljuaasen2003entanglement, osterloh2006enhancement, de2008symmetry}. These works investigated whether two-qubit concurrence can be modified or not in accordance with symmetry breaking for spin chain models. They specifically investigated the concurrence in relation to spin-rotation or spin-flip symmetry breaking while the Hamiltonians considered were translationally invariant. In our work, We focus on a Hamiltonian system in which translational symmetry breaking may occur and examine how entanglement is distributed among atoms at different locations in relation to phase transitions.

We specifically apply our method to an array of Rydberg atoms in 1D lattice that has attracted much attention as one of main platforms for quantum information processing due to its long coherence time and controllable interaction \cite{saffman2010quantum}. In particular, Rydberg atom arrays captured by optical tweezers scaled up in size and many significant experimental results were achieved in quantum simulation \cite{bernien2017probing, semeghini2021probing, ebadi2021quantum, scholl2021quantum, zhang2025probing} and quantum computing \cite{bluvstein2022quantum, bluvstein2024logical, reichardt2024logical}, {\it etc}.. The dynamics of Rydberg atoms is governed by the so-called Rydberg blockade, which prevents two neighboring atoms from occupying the Rydberg state simultaneously and thus provides a mechanism for rich phase diagrams in lattice systems breaking the translational symmetry \cite{fendley2004competing, samajdar2020complex, chepiga2021kibble, o2023entanglement, reinic2024finite}. Rydberg Hamiltonians also lead to exotic phenomena such as chiral transitions, floating phases \cite{samajdar2018numerical, chepiga2019floating, rader2019floating, yu2022fidelity, liao2024phase, garcia2024numerical}, and glassy or topological behavior \cite{verresen2021prediction, samajdar2023emergent, yan2023emergent}.

In this paper we study the entanglement structure of the 1-dim Rydberg atom lattice using density matrix renormalization group. We show how the analysis on the distribution of two-qubit concurrence over the entire system can provide useful information on quantum phases and their transitions in comparison with usual order parameters. This paper is structured as follows. In Sec. \ref{sec:preliminaries}, we briefly introduce the basic concepts and methods used in this work. In Sec. \ref{sec:1Drydberg} we discuss the symmetry breaking and the finite-size scaling of the Rydberg system in association with entanglement structure. We introduce an entanglement-based-structure factor $F_C(k)$ and compare it with the density order parameter $F_M(k)$ to investigate the characteristics of $\mathbb{Z}_n$ phase transitions in 1D Rydberg chain.  In Sec. \ref{sec:z4_comparison} we further investigate $\mathbb{Z}_4$ transitions along two different transition lines discussing the Ashkin-Teller and the incommensurate-commensurate transitions. In Sec. \ref{sec:measure_conc}, we propose how two-qubit concurrence can be measured using parametrized laser pulses in a Rydberg atom array. Finally, we summarize and discuss our results in Sec. \ref{sec:conclusion}.

\section{Preliminaries}\label{sec:preliminaries}
\subsection{Rydberg Hamiltonian}
In this section, we introduce the Rydberg Hamiltonian that has been intensively studied in the context of quantum simulation owing to the emergence of numerous exotic quantum phases with experimental controllability \cite{giudici2022dynamical} and scalability \cite{ebadi2021quantum, manetsch2024tweezer}. The basic components of Rydberg Hamiltonian are two internal energy levels, ground state $\ket{g}$ and Rydberg state $\ket{r}$. The Rydberg state is a highly excited state of an atom which strongly interacts with other atoms in  Rydberg states on $\mu\mathrm{m}$ length-scale. A Rydberg system with $N$ atoms is expressed as 
\begin{equation}\label{eq:rydberg_hamiltonian}
    H=\sum_{i=1}^N \bigg[\frac{\Omega}{2} (\ket{g_i}\bra{r_i}+\ket{r_i}\bra{g_i})-\Delta \hat{n}_i\bigg] + \sum_{i\neq j} \frac{V_{ij}}{2} \hat{n}_i \hat{n}_j.
\end{equation}
In Eq. (\ref{eq:rydberg_hamiltonian}), $\ket{g_i}$ and $\ket{r_i}$ are the internal states of the $i$th atom and $\hat{n}_i\equiv\ket{r_i}\bra{r_i}$ the occupation operator for the Rydberg state. The coefficients $\Omega$ and $\Delta$ denote the Rabi frequency of the system and the detuning of the corresponding laser, respectively. We see that a positive $\Delta$ favors the occupation of Rydberg state in each lattice site. The Rydberg interaction term, $V_{ij}=C_6/|\Vec{r}_i-\Vec{r}_j|^6$, depends on the relative distance and tends to forbid the simultaneous excitation of two adjacent atoms. That is, two atoms are blocked from being excited to the Rydberg state simultaneously if they are close enough. This effect may be quantitatively characterized by the radius of the Rydberg blockade $R_b$ defined by $V(R_b)=\Omega$. With $a$ the lattice constant, $R_b/a$ can be used as a dimensionless blockade radius. This Rydberg blockade enables a rich variety of quantum phases \cite{samajdar2020complex} and multi-qubit entangling gates for quantum information processing \cite{levine2019parallel}. For simplicity, we use $\Delta/\Omega$ as a parameter that indicates the preference for the Rydberg state of atoms, and $R_b/a$ as a parameter that represents the strength of the Rydberg interaction.

\subsection{Entanglement measure}
Entanglement is one of the fundamental features that distinguish quantum systems from classical ones. With the rise of quantum information theory, numerous works have focused on addressing quantum phase transitions via the amount of entanglement in a quantum system, usually quantified by the von Neumann entropy \cite{amico2008entanglement}. Remarkable results such as entanglement area laws \cite{hastings2007area,eisert2010colloquium} and the usefulness of matrix product states (MPS) \cite{schollwock2011density, orus2019tensor} were found to deepen our understanding of quantum many-body physics. In addition, several entanglement measures for mixed states were proposed to identify point-to-point or multi-qubit entanglement \cite{GUHNE20091}. 

To characterize entanglement structure for a mixed state of Rydberg system, we use concurrence \cite{wootters1998entanglement} between two qubits to quantify how entanglement between atoms is distributed in the Rydberg arrays of atoms. For a reduced density matrix $\rho$ of two qubits, we first obtain the spin-flipped state $\Tilde{\rho}$ as
\begin{equation}
    \Tilde{\rho}\equiv(\sigma_y \otimes \sigma_y) \rho^{*} (\sigma_y \otimes \sigma_y),
\end{equation}
where $\sigma_y$ and $\rho^{*}$ refer to the Pauli Y matrix and complex conjugation of $\rho$ in the computational basis, respectively. Then the concurrence $C(\rho)$ is given by 
\begin{equation}\label{eq:concurrence}
    C(\rho)\equiv\max\{0, \lambda_1-\lambda_2 -\lambda_3 - \lambda_4 \},
\end{equation}
where $\lambda_i$ ($i=1,2,3,4)$ are the eigenvalues of the Hermitian matrix $R\equiv \sqrt{\sqrt{\rho}\Tilde{\rho}\sqrt{\rho}}$ in decreasing order \cite{wootters1998entanglement}. 

On the other hand, if we intend to detect the entanglement between more than two qubits, we need a different approach due to the absence of an explicit formula for concurrence. We leverage negativity as a multi-qubit entanglement measure, which is a quantification based on partial transpose (PT) criterion \cite{vidal2002computable}. For an arbitrary density matrix $\rho$ in a bipartite Hilbert space $\mathcal{H}\equiv \mathcal{H}_A\otimes \mathcal{H}_B$, the negativity $\mathcal{N}(\rho)$ is defined as
\begin{equation}\label{eq:negativity}
    \mathcal{N}(\rho)\equiv\frac{||\rho^{T_A}||_1-1}{2},
\end{equation}
where $||\rho||_1\equiv \mathrm{tr}\sqrt{\rho^\dagger \rho}$ refers to the trace norm of the Hermitian operator $\rho$, and $\rho^{T_A}$ the partial transpose of $\rho$ with respect to the Hilbert space $\mathcal{H}_A$.

\subsection{Density matrix renormalization group}
Density matrix renormalization group (DMRG) is one of the most successful numerical methods to find the ground state of a many-body quantum system \cite{white1992density}. Combined with the matrix produc state (MPS) framework, DMRG can be understood as a successive procedure of discarding highly entangled states, which may negligibly affect the ground state of a complex quantum system \cite{verstraete2008matrix, schollwock2011density}. This procedure involves performing singular value decomposition and discarding singular values below a specified threshold.

In our study, we start DMRG from a small bond dimension and gradually increase it to find the ground state efficiently and reliably. We set the cutoff value from $10^{-7}$ to $10^{-12}$ according to the hardness of the problem. For example, multi-qubit negativity requires a higher accuracy than magnetization. We define the cutoff value $\epsilon$ by $\frac{\sum_{n\in \textrm{discarded}} \lambda_n^2 }{\sum_n \lambda_n^2}<\epsilon$, where $\lambda_n$ are singular values. We terminate the process if the relative decrease in the newly calculated energy is smaller than $10^{-11}$. For all calculations, we use the Julia package \texttt{ITensor.jl} \cite{fishman2022itensor}.

\begin{figure*}[t!]
    \includegraphics[width=1.0\linewidth]{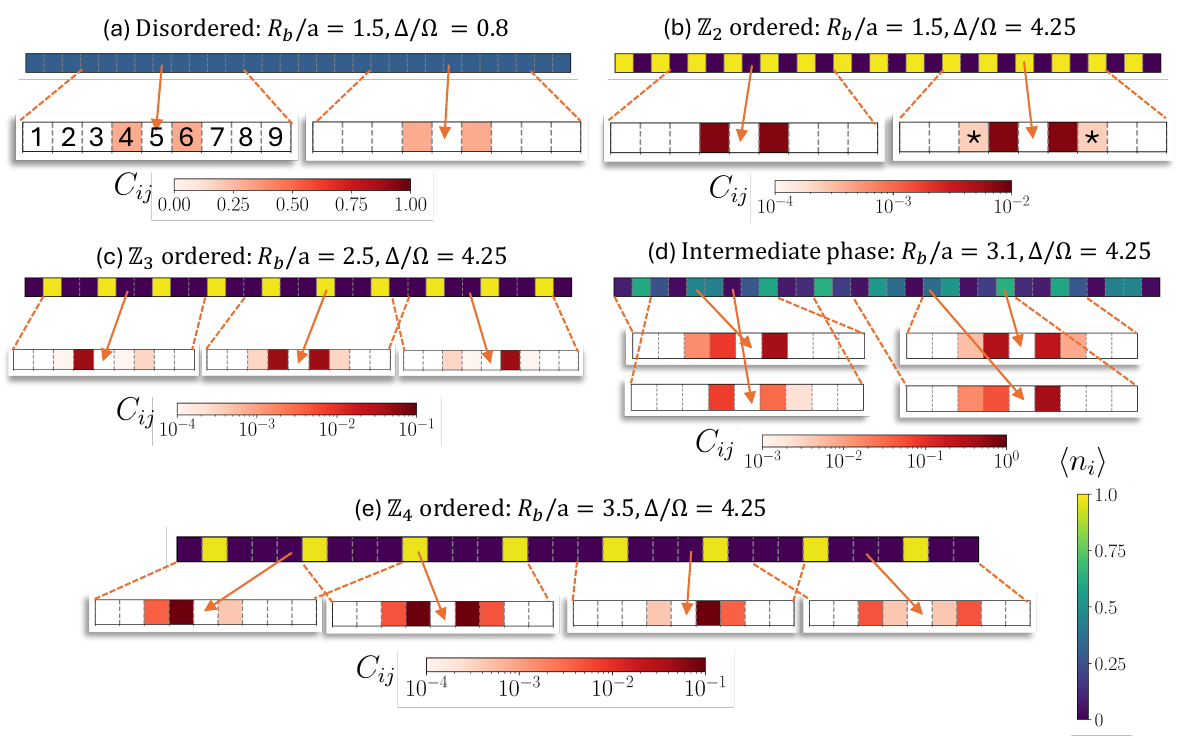} 
    \caption{Magnetization and concurrence profiles of representative phases on 1D Rydberg atom chain. In each part, the upper rectangle diagram represents the distribution of magnetization, i.e. the occupation of Rydberg atom $\langle n_i\rangle$ in lattice site $i$ with the color coding given at the right bottom. On the other hand, the lower rectangle diagrams represent the concurrence of the reference site with neighboring sites. Arrows from the magnetization profile to the concurrence profile indicate the reference site of two-qubit concurrence. In the concurrence profile, each colored box represents the strength of two-qubit concurrence with the reference atom. A white box means no entanglement with the reference site. In (a), a disordered-phase profile is shown. In (b), (c), and (e), the $\mathbb{Z}_2$, $\mathbb{Z}_3$, and $\mathbb{Z}_4$ ordered profiles are shown, respectively. In (d), the intermediate finite-size profile between the $\mathbb{Z}_3$ and $\mathbb{Z}_4$ ordered profiles is shown.}
    \label{fig:chain_profile}
\end{figure*}

\section{Entanglement order parameter in a 1D Rydberg chain}\label{sec:1Drydberg}
The Rydberg Hamiltonian in Eq. (\ref{eq:rydberg_hamiltonian}) offers a rich landscape of quantum phases even on a 1D chain. With a varying degree of Rydberg interaction, disordered  and ordered phases arise as a consequence of Rydberg blockade \cite{fendley2004competing, rader2019floating, chepiga2019floating, yu2022fidelity}. The crystalline ordered phases are predicted in the limit of $\Omega\rightarrow 0$, which is also related to the maximum independent set problem in combinatorial optimization \cite{pichler2018quantum, ebadi2022quantum}. For a small $R_b/a$, the effective interaction range is short and one atom can be excited to the Rydberg state among two adjacent atoms. This phase may be described as a density wave state of period 2, or simply $\mathbb{Z}_2$ ordered phase. As $R_b/a$ increases, the range of Rydberg blockade extends and the period of the Rydberg state also increases. We refer to the density wave state of period $n$ as $\mathbb{Z}_n$ ordered phase, in which $\mathbb{Z}_n$ translational symmetry is broken.

Together with their experimental realization \cite{bernien2017probing, keesling2019quantum}, the classification of phase transitions were substantially studied to date \cite{fendley2004competing, rader2019floating}. For example, it was shown that the disordered to ordered $\mathbb{Z}_2$ phase belongs to the (1+1)D Ising universality class with its well established scaling law \cite{sachdev1999quantum,sachdev2011quantum,yu2022fidelity}. However, things become more complicated for higher-period states. For disordered to $\mathbb{Z}_3$ ordered phase transition, three possibilities were suggested \cite{rader2019floating, yu2022fidelity}: phase transition belongs to (i) three-states Potts model, ii) chiral phase transition, and iii) incommensurate floating phase between them. Locating these three options on the phase diagram is not our main goal in this work, but we want to mention that extensive investigations on this long-standing topic in solid-state physics were made, e.g. to characterize commensurate to incommensurate phase transition  \cite{bak1982commensurate, ostlund1981incommensurate}. Ref. \cite{yu2022fidelity} proposed that the critical point of the three-state Potts model is surrounded by chiral phase transitions. As $R_b$ further increases, a floating phase emerges between the $\mathbb{Z}_3$ and $\mathbb{Z}_4$ phases, seamlessly connecting these two quantum phases. 

\subsection{Symmetry breaking in magnetization}
Symmetry breaking is one of the fundamental aspects of phase transitions in both classical and quantum regimes. It can be addressed by adopting appropriate order parameters that reflect the corresponding symmetries. For spin Hamiltonians including the Rydberg Hamiltonian in Eq. (\ref{eq:rydberg_hamiltonian}), one of the most conventional approach is to examine the distribution of magnetization. In this paper, we mainly use as a local observable the probability $\langle n_i \rangle$ that the $i$th atom is in the Rydberg state, which is also known as the Rydberg density. Using this distribution, we can identify how the translational symmetry is broken in 1D Rydberg chain as shown in the upper part of each plot in Fig. \ref{fig:chain_profile}.

In Fig. 1, the profiles of magnetization clearly show the transitions toward $\mathbb{Z}_2$, $\mathbb{Z}_3$ and $\mathbb{Z}_4$ phases with increasing $R_b$ at a fixed $\Delta/\Omega$. We also notice an intermediate profile between $\mathbb{Z}_3$ and $\mathbb{Z}_4$ phases in Fig. \ref{fig:chain_profile} (d). For the $\mathbb{Z}_4$ ordered phase, the transition character is also nontrivial. Previous works showed that the transition to the period-4 phase can be Ashkin-Teller-like on the commensurate line while the peak of the density structure factor may deviate from the period-4 wave vector in nearby parameter regimes \cite{chepiga2021kibble,maceira2022conformal}. We discuss these aspects in conjunction with our proposed entanglement characteristics later.

\subsection{Symmetry breaking in entanglement structure}
We now introduce our entanglement-structure-based approach and show how it characterizes quantum phases and their transitions. 
In contrast to the local observable i.e. magnetization, we look into the spatial structure of two-site concurrence between atoms at different positions along the lattice and identify the features relevant to translational symmetry breaking. 

To characterize the entanglement distribution under symmetry breaking, we evaluate the two-qubit concurrence in Eq. (\ref{eq:concurrence}) for every pair of atoms as depicted in the lower part of Fig. \ref{fig:chain_profile} (a)-(e). Since quantum states are repeated after $n$ atoms for an ideal $\mathbb{Z}_n$ ordered phase, we show the concurrence profiles with respect to $n$ different reference atoms in each phase. In Fig. \ref{fig:chain_profile}, the orange arrow from the magnetization profile to the concurrence profile indicates the position of the reference atom against which entanglement is evaluated with other atoms. The two-qubit concurrence between the reference atom and the other atoms is color-coded with a darker color representing a higher entanglement. For example, in Fig. 1 (a), the atom numbered 5 is the reference atom while the color in the box of atom 4 represents the degree of concurrence between atoms 4 and 5. White color means no entanglement so atom 5 has entanglement only with atoms 4 and 6, that is, the nearest neighbors.

For the $\mathbb{Z}_2$ ordered phase, the nearest-neighbor concurrence alone does not reflect the breaking of translational symmetry. Namely, in Fig. 1 (b), whether the reference atom is in a low-magnetization (left box diagram)  or high-magnetization state (right box diagram), the degree of concurrence with the nearst-neighbor atom is the same because it refers to the identical entanglement between the ground and the Rydberg-state atoms. However, we find that next-nearest-neighbor concurrence, which is marked with asterisk in Fig. 1 (b), exists only between two atoms with higher Rydberg densities $\langle n_i\rangle$, i.e. only between Rydberg-Rydberg atoms, not ground-ground atoms, thus reflecting the symmetry breaking of $\mathbb{Z}_2$-order . 

The profile in Fig. \ref{fig:chain_profile} (b) is evaluated at $\Delta/\Omega=4.25$, well inside the $\mathbb{Z}_2$ ordered region above a critical point $\Delta_c/\Omega$. There the next-nearest-neighbor concurrence between high-density $|r\rangle$ sites is visible. Near the transition point estimated from the magnetization structure $F_M(\pi)$, which will be shown later, this next-nearest-neighbor concurrence is much smaller. The $\mathbb{Z}_2$ case is more discussed separately in Appendix \ref{app:z2_entanglement}.

For $\mathbb{Z}_3$ and $\mathbb{Z}_4$ ordered phases, translational symmetry is also broken in the entanglement structure, as shown in Fig. \ref{fig:chain_profile} (c) and (e). It is worth noting that atoms with high Rydberg density $\langle n\rangle$ are generally more entangled with others than atoms with low Rydberg densities. In addition, we do not find two-qubit concurrence beyond next-next-nearest-neighbor pairs.
For the intermediate profile between $\mathbb{Z}_3$ and $\mathbb{Z}_4$ ordered phases in Fig. \ref{fig:chain_profile} (d), neither magnetization nor concurrence profile represents any period explicitly.

\begin{figure*}[t!]
    \includegraphics[width=1.0\linewidth]{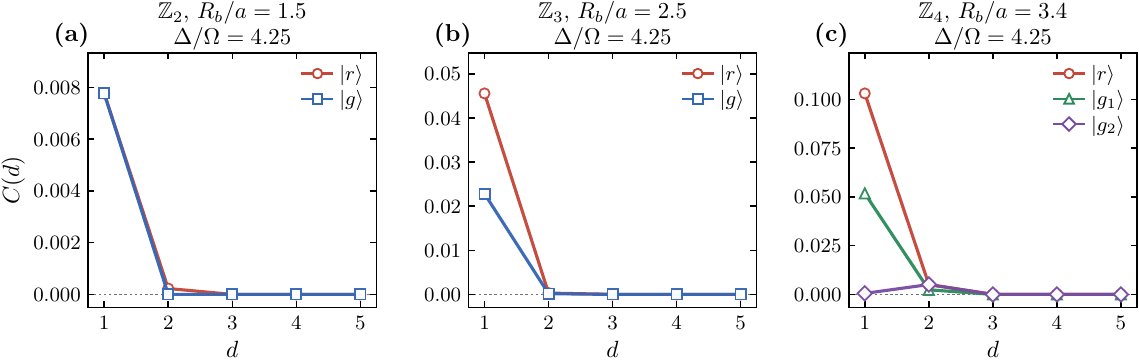}
    \caption{Site-resolved concurrence decay at $\Delta/\Omega=4.25$ for representative ordered profiles in 1D Rydberg chain. Panels show the $\mathbb{Z}_2$, $\mathbb{Z}_3$, and $\mathbb{Z}_4$ cases at the indicated values of $R_b/a$. Each curve corresponds to a different reference atom in a high-density (circle) or low-density (rectangle, rhombus) states, plotted as a function of distance from the reference atom. In (c), $|g_1\rangle$ and $|g_2\rangle$ denote the reference sites in low-density, nearest and next-nearest to a high-density $|r\rangle$ site, respectively.}
    \label{fig:concurrence_decay_site_resolved}
\end{figure*}

The concurrence profiles in Fig. \ref{fig:chain_profile} may be further characterized by plotting concurrence as a function of distance from each reference-site. Fig. \ref{fig:concurrence_decay_site_resolved} shows that concurrence is mostly concentrated within nearest-neighbor or next-nearest-neighbor atoms and that it rapidly decreases with distance. 

\subsection{Fourier analysis of the concurrence profile}
In order to identify period for each translational symmetry breaking, we can introduce a structure factor $F_M(k)$ based on the Fourier transform of Rydberg densities as
\begin{equation}\label{eq:ft_mag}
    F_M(k)= \frac{1}{N}\Bigg| \sum_{j=1}^N \langle n_j\rangle e^{-i jk}\Bigg|.
\end{equation}
According to its definition, $F_M(k)$ may show a pronounced peak at $k=2\pi/n$ for the $\mathbb{Z}_n$ ordered phase. For example, $F_M(\pi)$ can represent the so-called staggered magnetization, which detects antiferromagnetic structure.  

In our approach, we introduce a normalized Fourier amplitude of the local concurrence sum to characterize the spatial modulation of pairwise entanglement. We first define $C_j$, the total pairwise concurrence for the $j$th atom, as
\begin{equation}\label{eq:site_conc_sum}
   {C_j=\sum_{l\neq j} C_{j,l},}
\end{equation}
where $C_{j,l}$ refers to the concurrence between the $j$th and the $l$th atoms. Then we define
\begin{equation}\label{eq:ft_conc}
   {F_C(k)= \Bigg| \sum_{j=1}^N C_j e^{-i jk}\Bigg|/\sum_j C_j.}
\end{equation}
The normalization factor $\sum_j C_j$ in Eq. (\ref{eq:ft_conc}) is used to reduce the influence of the overall scale of the pairwise concurrence such that $F_C(k)$ mainly tracks the spatial modulation of the local concurrence sum. A comparison with a different normalization is given in Appendix \ref{app:fc_definition}.

\begin{figure*}[t!]
    \includegraphics[width=1.0\linewidth]{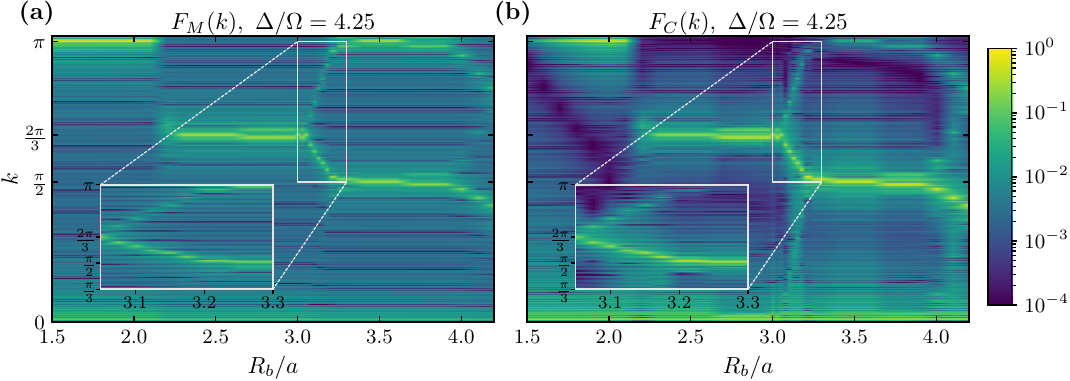} 
    \caption{Contour plots for Fourier transform (a) $F_M(k)$ of the Rydberg density and (b) $F_C(k)$ of the local concurrence sum, respectively, as functions of $R_b/a$ and $k$ at $\Delta/\Omega=4.25$. Bright ridges near $k=\pi$, $2\pi/3$, and $\pi/2$ mark the period-2, 3, and 4 density-wave regimes, respectively.}
    \label{fig:1D_ft}
\end{figure*}

We first show the value of $F_M(k)$ for different $R_b$ and $k$ in Fig. \ref{fig:1D_ft} (a). The bright lines at $k=\pi,2\pi/3$ and $\pi/2$ clearly indicate the locations of $\mathbb{Z}_2$, $\mathbb{Z}_3$ and $\mathbb{Z}_4$ ordered phases, respectively. Interestingly, there are additional bright structures (boxed in figures) between $\mathbb{Z}_3$ and $\mathbb{Z}_4$. Ref. \cite{rader2019floating} mentioned that floating phases are located between $\mathbb{Z}_3$ and $\mathbb{Z}_4$ ordered phases and that the density structure factor evolves continuously between these ordered phases in infinite-DMRG calculations. We do not discuss this issue in detail but our finite-DMRG approach gives a similar feature.

In comparison, we plot our new structure factor $F_C(k)$ based on entanglement in the same parameter range in Fig. \ref{fig:1D_ft} (b). The bright ridges in $F_C(k)$ occur near the same period-selective Fourier components as in $F_M(k)$, indicating that the spatial modulation of pairwise entanglement carries the period information in the ordered regimes.

\subsection{Finite-size scaling of $\mathbb{Z}_n$ ordered phases}

\begin{figure*}[t!]
    \includegraphics[width=1.0\linewidth]{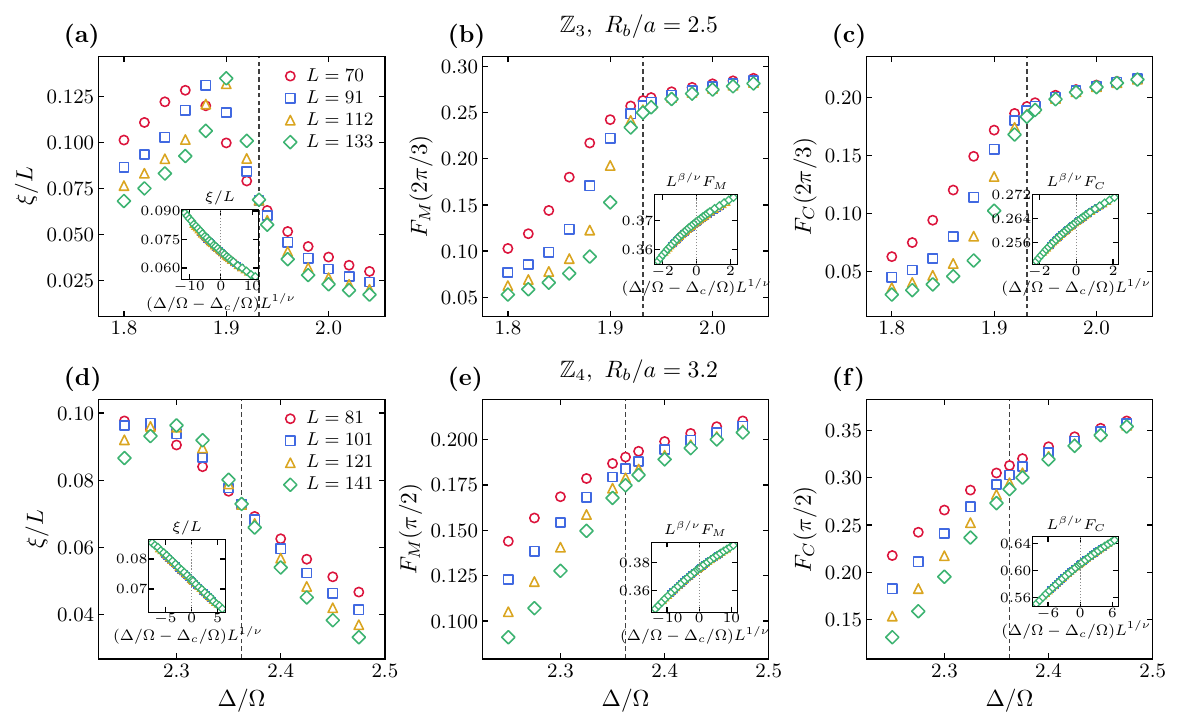} 
    \caption{Finite-size scaling for $\mathbb{Z}_3$ and $\mathbb{Z}_4$ ordered phases. In (a)-(c), the $\mathbb{Z}_3$ transition for $R_b/a=2.5$ is analyzed with the correlation length $\xi/L$, the magnetization structure factor $F_M(k)$ and the entanglement structure factor $F_C(k)$ at $k=\frac{2\pi}{3}$. In (d)-(f), the $\mathbb{Z}_4$ transition for $R_b/a=3.2$ and is analyzed at $k=\frac{2\pi}{4}$.}
    \label{fig:z3_z4_scaling}
\end{figure*}

In this subsection we conduct finite-size scaling analyses for $\mathbb{Z}_3$ and $\mathbb{Z}_4$ ordered phases (Fig. \ref{fig:z3_z4_scaling}). To begin with, we adopt as another indicator of phase transition a dimensionless correlation length $\xi/L$, where $\xi$ is the second-moment correlation length estimated from the density structure factor and $L$ the lattice size. We first define a density-density correlation,
\begin{equation}\label{eq:connected_density_corr}
    G_{ij}=\langle n_i n_j\rangle-\langle n_i\rangle\langle n_j\rangle,
    \end{equation}
and define the density structure factor as
\begin{equation}\label{eq:density_structure_factor}
    S(k)=\frac{1}{L}\sum_{i,j}G_{ij}\cos[k(i-j)].
\end{equation}
Following the standard second-moment estimator based on the structure factor \cite{sandvik2010computational}, we obtain the correlation length as
\begin{equation}\label{eq:second_moment_xi}
    \xi =
    \frac{1}{2\sin(\delta k/2)}
    \left[
    \frac{S(k^\ast)}
    {\{S(k^\ast+\delta k)+S(k^\ast-\delta k)\}/2}
    -1
    \right]^{1/2},
\end{equation}
where $k^\ast=2\pi/n$ is the period-$n$ ordering wave vector, $\delta k=2\pi/L$, and the average over $k^\ast\pm\delta k$ is used to reduce the left-right asymmetry of the finite open chain.
Because $\xi/L$ is dimensionless, its finite-size scaling collapse is obtained by rescaling only the distance from the transition point, without an additional power of $L$
\cite{fisher1972scaling,binder1981finite,nohadani2005quantum,wierschem2014characterizing}. That is, we use the form
  \begin{equation}\label{eq:ansatz_xi}
      \frac{\xi}{L}
      =
      f_{\xi}\!\left(
      L^{1/\nu}\left[(\Delta/\Omega)-(\Delta/\Omega)_c\right]
      \right)
  \end{equation}
  to find the correlation-length exponent $\nu$ with $(\Delta/\Omega)_c$ the at transition point.

On the other hand, the Fourier transform of magnetization as a conventional order parameter also follows a scaling law at phase transition. We introduce an ansatz of the scaling behavior for $F_M$ as
\begin{equation}\label{eq:ansatz_F_M}
    F_M = L^{-\beta/\nu}f_{F_M}(L^{1/\nu}[(\Delta/\Omega)-(\Delta/\Omega)_c]),
\end{equation}
where $\nu$ and $\beta$ are critical exponents \cite{fisher1972scaling,nohadani2005quantum, wierschem2014characterizing, bhandari2017first} and $(\Delta/\Omega)_c$ the critical detuning at transition point.

In parallel we also adopt a finite-size scaling for our concurrence structure $F_C$ as
\begin{equation}\label{eq:ansatz_F_C}
    F_C = L^{-\beta/\nu}\mathcal{F}_{F_C}(L^{1/\nu}[(\Delta/\Omega)-(\Delta/\Omega)_c]).
\end{equation}
We test the introduced finite-size scaling forms numerically. Additionally, we further compare the behaviors of  $F_C$ with individual concurrences in Appendix \ref{app:fc_definition} . In these analyses, an individual concurrence with respect to a single reference site does not provide a robust transition marker unlike the Fourier sum $F_C(k)$ encompassing entanglement over the entire lattice shown in Fig. 4.
The representative scans of individual concurrence values are shown in Fig. \ref{fig:app_individual_concurrence_scan}. Thus, the critical behavior, or universality if any, around phase transitions can be found only through the analysis of the overall entanglement structure in connection to translational symmetry breaking as shown below.

For the $\mathbb{Z}_3$ ordered phase, we plot $\xi/L$, $F_M(2\pi/3)$ and $F_C(2\pi/3)$ in Fig. \ref{fig:z3_z4_scaling} (a)-(c). We find that the curves of $\xi/L$ at different lattice sizes $L$ cross almost identically at $(\Delta/\Omega)_c=1.9322(5)$ in Fig. \ref{fig:z3_z4_scaling} (a). The inset also shows the collapse of the finite scaling to a single value of exponent $\nu\simeq 0.676(2)$. The uncertainty in the critical point is estimated from the stability under the fit-window and system-size subset choices. 
In Fig. \ref{fig:z3_z4_scaling} (b), the slope of $F_M(2\pi/3)$ near the same critical point becomes steeper as the system size increases. The inset there shows the collapse to $\nu\simeq 0.870(24)$ and $\beta/\nu\simeq 0.07988(6)$. In Fig. \ref{fig:z3_z4_scaling}(c), $F_C(2\pi/3)$ shows a similar behavior near the transition and its inset gives a stable collapse to $\nu\simeq 0.875(9)$ and $\beta/\nu\simeq 0.07307(6)$ close to those values in the magnetization $F_M(2\pi/3)$. 
The conformal period-3 transition in Rydberg chains is expected to be in the three-state Potts universality class, where $\nu=5/6$ and $\beta/\nu=2/15$ \cite{maceira2022conformal}. While our finite-size scaling does not precisely indicate this universality class, the fitted exponents from both of $F_M(2\pi/3)$ and $F_C(2\pi/3)$ are very close to each other and both values are also comparable with the three-state Potts benchmark.

We extend our investigation to $\mathbb{Z}_4$ ordered phase ($k=\pi/2$) as shown in Fig. \ref{fig:z3_z4_scaling} (d)-(f). In Fig. \ref{fig:z3_z4_scaling} (d), $\xi/L$ gives a crossing and a collapse to $(\Delta/\Omega)_c=2.3629(6)$ and $\nu\simeq 0.850(6)$.
In Fig. \ref{fig:z3_z4_scaling} (e) and (f), both $F_M(\pi/2)$ and $F_C(\pi/2)$ increase through the same transition region. Their insets show finite-size collapse to the values $\nu\simeq 0.781(2)$, $\beta/\nu\simeq 0.15279(9)$ for $F_M$, and $\nu\simeq 0.848(2)$, $\beta/\nu\simeq 0.1496(2)$ for $F_C$. The exponent obtained from $F_C(\pi/2)$ is close to the value from $\xi/L$ and also to the Ashkin-Teller value $\nu\simeq 0.80$ reported for the conformal period-4 transition in Rydberg chains \cite{maceira2022conformal}.

For the density order parameter $F_M$ on the Ashkin-Teller cut, Ref. \cite{maceira2022conformal} expects $\beta=\nu/8$, or $\beta/\nu=1/8$. Our fitted $\beta/\nu$ values are larger for both $F_M$ and $F_C$ than this value. Nevertheless, the finite-size behaviors of $F_M$ and $F_C$ are close to each other for the $\mathbb{Z}_4$ transition. In the next section, we compare two period-4 cuts with different behaviors in $S(k)$ and entanglement entropy.

\section{Comparison of $\mathbb{Z}_4$ phase transitions}\label{sec:z4_comparison}
\begin{figure*}[t!]
    \includegraphics[width=1.0\linewidth]{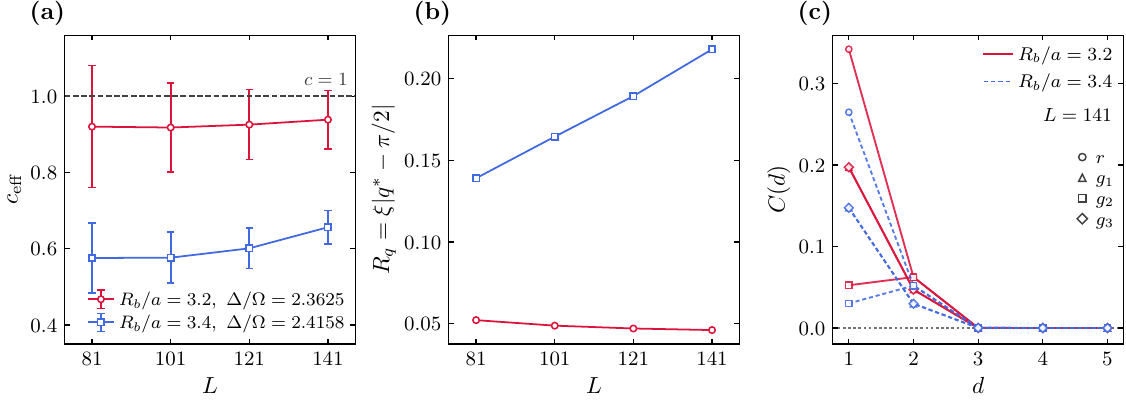} 
    \caption{Comparison of two cuts into the period-4 ordered phase. In (a), $c_{\mathrm{eff}}$ is extracted from entanglement entropy profiles at $\Delta/\Omega=2.3625$, close to the transition point $(\Delta/\Omega)_c=2.3629(6)$ obtained from $\xi/L$ for $R_b/a=3.2$, and at $\Delta/\Omega=2.4158$ for $R_b/a=3.4$. In (b), $R_q=\xi|q^\ast-\pi/2|$ is plotted, where $q^\ast$ is the peak position of $S(k)$. In (c), site-resolved concurrence decay is plotted as a function of distance $d$ for $L=141$ at the same values of $\Delta/\Omega$.}
    \label{fig:z4_transition_comparison}
\end{figure*}

The $\mathbb{Z}_4$ ordered phase provides another interesting test because the transition characteristics to the same period-4 density wave can depend on the interaction strength. On the commensurate transition line, previous works showed that the period-4 transition of Rydberg chains is described by the Ashkin-Teller universality class with $\nu\simeq 0.80$ \cite{maceira2022conformal}. Let $q^\ast$ denote the wave vector at which the density structure factor $S(k)$ is maximal, e.g. $q^\ast=\pi/2$ for period-4. In nearby parameter regimes, the peak of $S(k)$ may deviate from $\pi/2$ on the disordered side, a behavior discussed in terms of chiral transition or floating phase \cite{chepiga2021kibble,maceira2022conformal}. In this section, we compare the two transition cuts at $R_b/a=3.2$ and $R_b/a=3.4$ representing the two transition lines mentioned above and see which features are visible in the density order parameter and the concurrence. Specifically, in Fig. \ref{fig:z4_transition_comparison} (a), we use the entanglement entropy profile near the transition point to extract the effective central charge. In Fig. \ref{fig:z4_transition_comparison} (b), we plot $R_q=\xi|q^\ast-\pi/2|$, which measures the deviation of the peak of $S(k)$ from the period-4 wave vector. In Fig. \ref{fig:z4_transition_comparison} (c), we compare the site-resolved concurrence decay at the same values of $\Delta/\Omega$. 

For the entropy analysis, we fit the open-boundary entanglement entropy profile to
\begin{equation}\label{eq:ceff_fit}
S(l)=\frac{c_{\mathrm{eff}}}{6}
\log\!\left[\frac{2L}{\pi}\sin\!\left(\frac{\pi l}{L}\right)\right]+s_0,
\end{equation}
after excluding bonds close to the open edges. The fitted coefficient $c_{\mathrm{eff}}$ is used as a finite-size estimate of the central charge on cuts where conformal scaling is expected \cite{calabrese2004entanglement,calabrese2009entanglement}.

(i) $R_b/a=3.2$ : The collapse of $\xi/L$  gives $(\Delta/\Omega)_c=2.3629(6)$ and we evaluate the entropy profile closely at $\Delta/\Omega=2.3625$ in Fig. \ref{fig:z4_transition_comparison} (a). This is the Ashkin-Teller-like period-4 cut used in Fig. \ref{fig:z3_z4_scaling} (d)-(f). In Fig. \ref{fig:z4_transition_comparison} (a), the extracted $c_{\mathrm{eff}}$ stays close to $c=1$ as the system size increases. In Fig. \ref{fig:z4_transition_comparison} (b), $R_q$ slightly decreases with system size indicating $q^\ast$ staying close to $\pi/2$. These results are consistent with the Ashkin-Teller-like period-4 transition. Along the same cut, $\xi/L$, $F_M(\pi/2)$, and $F_C(\pi/2)$ also change consistently with the transition as shown in Fig. \ref{fig:z3_z4_scaling} (d)-(f).

(ii) $R_b/a=3.4$: For $\Delta/\Omega=2.4158$, the ordered side is still a period-4 density wave and $F_M(\pi/2)$ and $F_C(\pi/2)$ grow as the ordered phase is approached. However, Fig. \ref{fig:z4_transition_comparison} (a) does not show the same $c_{\mathrm{eff}}\simeq 1$ behavior as the one given at the $R_b/a=3.2$ cut. In Fig. \ref{fig:z4_transition_comparison} (b), $R_q$ also increases with system size indicating that the peak position $q^\ast$ of $S(k)$ is not fixed at $\pi/2$ on the disordered side. Thus, this cut is different from the Ashkin-Teller-like transition, even though the ordered phase reached after the transition has the same period-4 structure. Detailed raw curves and collapse plots for the two $\mathbb{Z}_4$ cuts are given in Appendix \ref{app:z4_scaling_diagnostics}.

The concurrence decays shown in Fig. \ref{fig:z4_transition_comparison} (c) should be interpreted together with this distinction. In both cuts, $F_C(\pi/2)$ displays the period-4 modulation of the local concurrence sum. The two different cuts are thus distinguished more clearly by the entropy behavior and $R_q$ in Fig. \ref{fig:z4_transition_comparison} (a) and (b), with more details in Appendix \ref{app:z4_wavevector}.

\section{Measuring concurrence through parametrized laser pulses}\label{sec:measure_conc}
\begin{figure*}[t!]
    \includegraphics[width=1.0\linewidth]{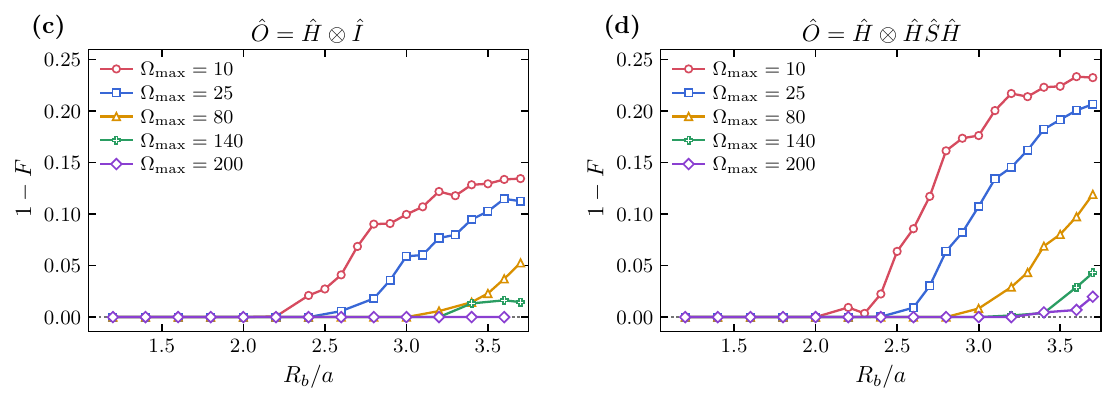} 
    \caption{Infidelity $1-F$ for target two-qubit rotations by optimized pulse implementation . The target measurement is (a) $X\otimes Z$ corresponding to  the unitary operation $U_{\mathrm{tar}}=H\otimes I$ and (b) $X\otimes Y$, corresponding to $U_{\mathrm{tar}}=H\otimes HSH$. $H$ is the Hadamard gate and $S=\mathrm{diag}(1,i)$ the single-qubit phase gate. The infidelity is shown as a function of the Rydberg-interaction strength $R_b/a$ at a different  $\Omega_{\max}/\Omega_0$. The local detuning bound is fixed at $\Delta_{\max}=20\Omega_0$.}
    \label{fig:local_rotation}
\end{figure*}

Measuring two-qubit concurrence is straightforward in conventional digital quantum hardware, where it can be obtained e.g. from quantum state tomography. Two-qubit state tomography can be implemented by measuring two target qubits in the basis of $\sigma_i\sigma_j$, where $\sigma_i\in\{I,X,Y,Z\}$. On the other hand, a different approach is required in analog quantum hardware. In our case, the Rydberg system may evolve according to the Rydberg Hamiltonian in Eq. (\ref{eq:rydberg_hamiltonian}). The problem is that the Rydberg interaction $\sum_{i\neq j}V_{ij}\hat{n}_i \hat{n}_j$ in the Hamiltonian prevents independent single-qubit rotations, which is essential for quantum state tomography. There were studies to realize digital quantum computation only through Hamiltonian evolution, however, Under special conditions \cite{cesa2023universal,votto2024universal}. 
As our task is to implement independent basis rotations only for the desired two qubits, we present a simpler route for this problem. 

One direct approach is to abruptly increase the laser intensity $\Omega$ to minimize the effect of the Rydberg interaction. In Ref. \cite{semeghini2021probing}, they succeeded in applying single-qubit rotations for independent triangle groups on a kagome lattice in this way. However, quantum state tomography requires more freedom in the rotation angle. Another way to implement single-qubit rotations is to parametrize laser pulses. Ref. \cite{chevallier2024variational} showed that arbitrary single-qubit rotations up to 8 qubits can be implemented with a very high fidelity on 1D chain by adjusting the shape of laser pulses. We note that this scheme includes a global rotation part whose feasibility in a larger lattice has not been proven.

In this section, we focus on the fact that we do not have to rotate every single qubit in our case. For example, measuring the selected pair in the $X\otimes Z$ basis requires a basis rotation on the selected pair before the final $Z$-basis readout. Since two-qubit concurrence only requires the reduced density matrix of the selected pair, atoms outside the selected pair can be eliminated before the tomography rotation. We apply this site-selective elimination on non-target sites after the many-body state is prepared. Ground-state atoms can be removed by resonant push-out light while Rydberg-state atoms can be removed via loss by applying the anti-trapping tweezer potential on the non-target sites. Site-selective push-out and mid-circuit readout were demonstrated in reconfigurable neutral-atom arrays \cite{bluvstein2024logical}, and the use of tweezer ramp-up to eject Rydberg-state atoms in a ground-Rydberg encoding was reported in Rydberg tweezer arrays \cite{anand2024dual}. Because the erasure outcomes are not used for data analysis, this step implements the partial trace over the non-target subsystem before the tomography pulse.

The same erasure step must not be applied to the target sites before the tomography pulse. In a ground-Rydberg encoding, restoring the tweezer trap on a target atom would eject its Rydberg component and would therefore constitute a readout operation. The erasure step should also be completed before the global Rabi pulse and should be fast compared with the Rydberg lifetime and with the target--non-target interaction time scale. Once the non-target atoms are excluded, the global Rabi beam acts only on the selected pair, so it can be used for the optimized two-atom basis rotation. The pulse fidelities reported below do not include the erasure error itself.

The remaining tomography is then reduced to implementing basis rotations on two target atoms. For a two-qubit reduced density matrix, the required measurements are the combinations of Pauli bases such as $X\otimes Z$ and $X\otimes Y$. In an analog Rydberg system, these basis rotations are not native independent gates because the two atoms still evolve under the Rydberg Hamiltonian during the pulse. We therefore optimize the global Rabi frequency and the local detuning so that the two-atom Hamiltonian evolution approximates the desired basis rotation. This construction is related to the basis-rotation measurement used in Rydberg-array experiments. For example, Ref. \cite{semeghini2021probing} used Rydberg Hamiltonian evolution before the final $Z$-basis readout to access non-$Z$ correlators. Local light shifts provide another possible route to tune selected atoms in and out of resonance with a global Rydberg drive \cite{burgers2022controlling}, and local detuning is also available as a control knob in programmable neutral-atom hardware \cite{wurtz2023aquila}.

A figure of merit to measure a successful realization of gate operation is the fidelity $F$ between our designed Hamiltonian evolution and the target operator $O$, which is defined as
\begin{equation}
    F = \frac{1}{\text{dim(O)}}\Bigg|\text{tr}\Bigg[O^\dagger \prod_{j=1}^{2^p}\exp\bigg(-i\frac{\Delta t}{\hbar}H(\theta_j)\bigg)\Bigg]\Bigg|.
\end{equation}
We numerically find that parametrized global Rabi frequency $\Omega$ and local detunings $\Delta_i$ in an experimentally accessible regime produce arbitrary single-qubit rotations for two qubits in high fidelity. In Fig. \ref{fig:local_rotation}, we plot the infidelity $1-F$ according to the interaction strength and the pulse-amplitude bound. For each value of $R_b/a$, we normalize the two-atom Hamiltonian by $\Omega_0=C_6/[(R_b/a)]^6$. Therefore, plots are given in terms of the dimensionless bounds $\Omega_{\max}/\Omega_0$ and $\Delta_{\max}/\Omega_0$. In these figures, $\Delta_{\max}=20\Omega_0$ is fixed and $\Omega_{\max}$ is varied. 

For the $X\otimes Z$ measurement, we need to implement a Hadamard gate and a null operation, i.e. $U_{tar}=H\otimes I$ onto two target qubits. The numerical study shows that the infidelity remains below $10^{-2}$ up to a higher Rydberg interaction strength with a larger Rabi frequency, for instance, up to $R_b/a\simeq 3.2$ with $\Omega_{\max}=80\Omega_0$, up to $R_b/a\simeq 3.5$ with $\Omega_{\max}=140\Omega_0$, and over the full plotted range with $\Omega_{\max}=200\Omega_0$. On the other hand, the $X\otimes Y$ measurement is harder because both target atoms require nontrivial rotations: We need to implement  $U_{tar}=H\otimes HSH$, where $S=\mathrm{diag}(1,i)=e^{i\pi/4}R_z(\pi/2)$ is a phase gate. In this case, the infidelity remains below $10^{-2}$ up to $R_b/a\simeq 3.0$ with $\Omega_{\max}=80\Omega_0$, up to $R_b/a\simeq 3.4$ with $\Omega_{\max}=140\Omega_0$, and up to $R_b/a\simeq 3.6$ with $\Omega_{\max}=200\Omega_0$. 
It thus seems feasible to accomplish a high-fidelity tomography to measure concurrence once an intense Rabi pulse is applied together with the elimination of non-target atoms.

\section{Conclusion}\label{sec:conclusion}

Phase transitions are usually analyzed using order parameters based on local physical quantities like magnetization to characterize symmetry breaking. In quantum domain, quantum fluctuations particularly in the form of quantum correlations play a crucial role in shaping quantum phases that emerge under multi-body interactions. Thus, to deepen our understanding of quantum many-body physics, it is of great importance to develop a framework identifying the role and the characteristics of quantum entanglement in quantum phase transitions.
In this work we have proposed the entanglement structure along the lattice as a theoretical and experimental tool to investigate quantum phases and their transitions in 1D Rydberg atom array. 

We specifically used two-qubit concurrence as a measure of local entanglement and suggested to examine how the sum of pairwise entanglement evaluated at each site is distributed over the entire system. For this purpose, we introduced the Fourier analysis $F_C(k)$ of local entanglement-sum which turns out to characterize well quantum phase transitions in 1D Rydberg array. 
We numerically calculated two-qubit concurrence through DMRG and compared its spatial structure with the Rydberg density profile. In $\mathbb{Z}_3$ and $\mathbb{Z}_4$ ordered phases, the concurrence profile exhibits the same local period and the characteristics as the magnetization density profile.
The finite-size scaling of $F_C(k)$ near the $\mathbb{Z}_3$ and $\mathbb{Z}_4$ transition cuts also arise similar to the correlation length $\xi/L$ and the density order parameter $F_M(k)$.

Our method adopts a nonlocal property of the system while the conventional order parameter based on magnetization is a local quantity. It is thus interesting to note that both local and nonlocal approaches address quantum phase transitions equally well in most of the parameter regions. On the other hand, there exist some cases, e.g. $\mathbb{Z}_2$-phase transition due to next-nearest-neighbor concurrence, where disparity arises between two approaches. Whether this is a pathological phenomenon is worth further investigation in future. Namely, it will be important, and nontrivial, to know whether  local and nonlocal methods should generally agree in characterizing quantum phase transitions. In a related context, our analysis of $\mathbb{Z}_4$-order along two different transition cuts showed that $F_C(\pi/2)$ captures the period-4 local concurrence modulation for both transition regimes, whereas the two cuts are distinguished more clearly by the peak position of the density structure $S(k)$ and by entanglement-entropy scaling. 

Finally, we proposed how two-qubit concurrence can be measured by optimized Hamiltonian evolution using currently available neutral-atom control techniques. Since concurrence only requires the reduced density matrix of a selected pair, atoms outside the selected pair can be erased before the tomography rotation. After this site-selective erasure step, parametrized global Rabi and local detuning pulses can implement the required two-atom basis rotations with high target-pair fidelity in the parameter range considered here. We hope that our methodology for investigating quantum phase transitions can provide a new perspective on the connection between quantum information and quantum many-body physics to stimulate further related works.

\section{Acknowledgment}

This work was supported by the National Research Foundation of Korea (NRF) grants funded by the MSIT (Grant No. RS-2025-00515537), the Institute for Information \& Communications Technology Promotion (IITP) grants funded by the Korean government (MSIP) (Grant Nos. RS-2019-II190003 and RS-2025-02304540), the National Research Council of Science \& Technology (NST) (Grant No. GTL25011-401), and the Korea Institute of Science and Technology Information (KISTI) (Grant No. P26028).

\appendix

\begin{figure*}[t!]
    \includegraphics[width=1.0\linewidth]{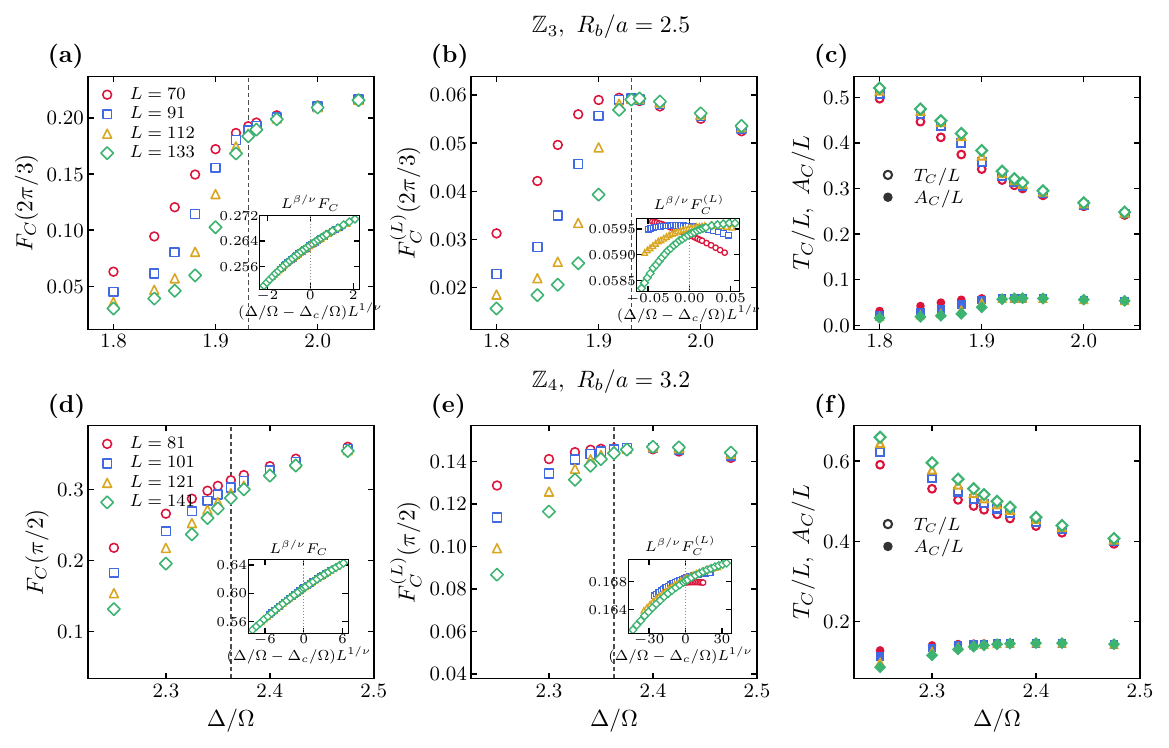}
    \caption{Comparison of  $F_C(k)=A_C(k)/T_C$ in (a) and (d) and $F_C^{(L)}(k)=A_C(k)/L$ in (b) and (e) to characterize phase transitions. (a)-(c) $\mathbb{Z}_3$ cut at $R_b/a=2.5$ and (d)-(f)  $\mathbb{Z}_4$ cut at $R_b/a=3.2$. The insets show the corresponding collapses in the same $\Delta/\Omega$- window after fixing the critical value $(\Delta/\Omega)_c$ at the one obtained from the analysis of $\xi/L$.  (c) and (f) show $T_C/L$ and $A_C/L$, which separate the total extracted concurrence scale from its Fourier amplitude.}
    \label{fig:app_fc_definition_comparison}
\end{figure*}

\newpage

\section{Defining $F_C$ with different  normalizations}\label{app:fc_definition}
In the main text, we evaluate the local concurrence sum at site $i$ as $C_i=\sum_{j\ne i}C_{ij}$ and define the entanglement structure factor as $F_C(k)=A_C(k)/T_C$, where $A_C(k)=|\sum_j C_j e^{-ikj}|$ and $T_C=\sum_j C_j$. This denominator removes the dependence on the total concurrence, so $F_C(k)$ is intended as a normalized measure of the spatial modulation in entanglement.

To test the effect of this denominator choice, we compare it with $F_C^{(L)}(k)=A_C(k)/L$ normalized with lattice size $L$. The two definitions are evaluated using the same concurrence data as a function of $\Delta/\Omega$ in Fig. \ref{fig:app_fc_definition_comparison}. In performing a finite-scaling analysis for both definitions, the transition point $(\Delta/\Omega)_c$ is fixed at the one obtained from the analysis of the correlation length $\xi/L$. Thus, only the exponents $\nu$ and $\beta/\nu$ are individually evaluated in each scaling.

Fig. \ref{fig:app_fc_definition_comparison} shows why we adopt $A_C/T_C$ as the entanglement structure factor in the main text. In Fig. \ref{fig:app_fc_definition_comparison} (a) and (d), $F_C(k)$ gives a better-resolved collapse for the parameter cuts studied here. In Fig. \ref{fig:app_fc_definition_comparison} (b) and (e), the lattice-size-normalized $F_C^{(L)}(k)$ is constructed from the same Fourier amplitudes but appears to carry a part of smooth variation  of the total concurrence. We therefore regard $A_C/T_C$ as a more sensible definition to focus on the spatial pattern of the concurrence sum rather than on the total amount of extracted two-site concurrence. The lattice-size-normalized  $A_C(k)/L$ mixes two effects: the smooth change in the total  pairwise concurrence and the growth of its spatial modulation. The ratio $A_C(k)/T_C$ suppresses the former and is thus better suited for comparing the modulation pattern across system sizes.

\begin{figure*}[tbh!]
    \includegraphics[width=1.0\linewidth]{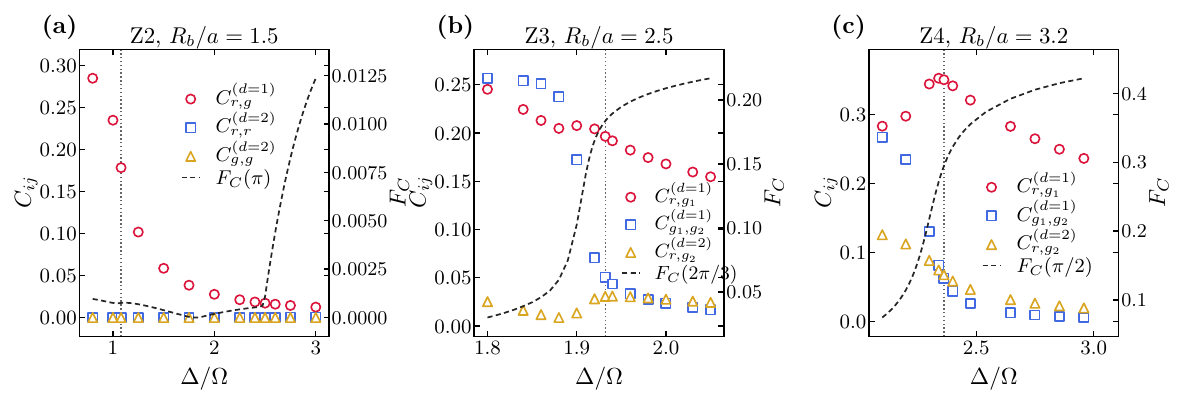}
    \caption{Representative individual concurrences between two single atoms as a function of $\Delta/\Omega$. The colored points show selected local concurrence channels, and the dashed black curves show the corresponding $F_C(k)$ with values on the right axis. In (a), the $\mathbb{Z}_2$ transition at $R_b/a=1.5$ is shown with $k=\pi$. In this case $C_{r,r}^{(d=2)}$ (entanglement between next-nearest atoms in Rydberg states) becomes nonzero only after $\Delta/\Omega\simeq 2.5$ and remains small with a maximum of order $3\times10^{-4}$ in the plotted range, while $C_{g,g}^{(d=2)}$ stays zero. In (b), the $\mathbb{Z}_3$ transition at $R_b/a=2.5$ is shown with $k=2\pi/3$. In (c), the $\mathbb{Z}_4$ transition at $R_b/a=3.2$ is shown with $k=\pi/2$.}
    \label{fig:app_individual_concurrence_scan}
\end{figure*}

For comparison we also analyze representative individual concurrences at single sites as a function of $\Delta/\Omega$. In Fig. \ref{fig:app_individual_concurrence_scan}, the colored points show selected local concurrence channels while the dashed black curve shows the corresponding $F_C(k)$ on the right axis. 
In Fig. \ref{fig:app_individual_concurrence_scan} (a) representing $\mathbb{Z}_2$ transition, we observe a peculiar feature of pairwise entanglement unlike $\mathbb{Z}_3$ and $\mathbb{Z}_4$ transitions: The local next-nearest-neighbor concurrence between the high-density sites turns on substantially after the transition point estimated from $F_M(\pi)$. Fig. \ref{fig:app_individual_concurrence_scan} (b) and (c) also show that individual concurrences at single sites do not manifest the transition behavior appearing in the collective quantity $F_C(k)$. These results indicate the importance of looking into the overall entanglement structure over the entire lattice, not individual ones, in characterizing quantum phases and their transitions.

\section{Detailed finite-size scaling curves for $\mathbb{Z}_4$ transition}\label{app:z4_scaling_diagnostics}

In Sec. \ref{sec:z4_comparison}, we compare two transition cuts into the period-4 ordered phase. Here we show the corresponding finite-scaling curves for the fixed period-4 components of $F_M(\pi/2)$ and $F_C(\pi/2)$, respectively. The figures here are meant to show in more details how the Fourier amplitudes used in the main text behave around the transition point.

\begin{figure*}[tbh!]
\includegraphics[width=1.0\linewidth]{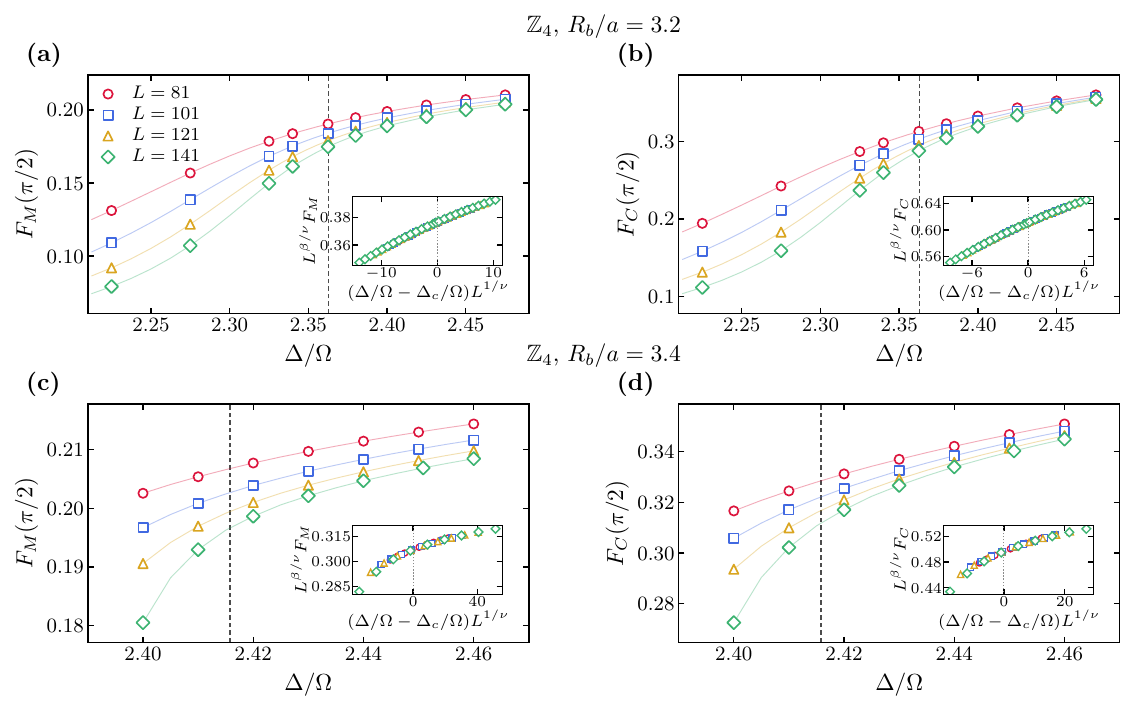}
\caption{Detailed $\mathbb{Z}_4$ finite-size scaling curves for $F_M(\pi/2)$ and $F_C(\pi/2)$. (a), (b) $R_b/a=3.2$, and (c), (d)  $R_b/a=3.4$, with $(\Delta/\Omega)_c$ fixed at 2.4158. The exponent values are quoted in the text.}
\label{fig:app_z4_scaling_diagnostics}
\end{figure*}

For $R_b/a=3.2$, the data in Fig. \ref{fig:app_z4_scaling_diagnostics} (a) and (b) are the same data used in Fig. \ref{fig:z3_z4_scaling} (e) and (f), in the same range of $\Delta/\Omega$ with $(\Delta/\Omega)_c=2.3629(6)$. The collapse with the fixed $(\Delta/\Omega)_c$ gives $\nu\simeq0.772(10)$, $\beta/\nu\simeq0.155(2)$ for $F_M(\pi/2)$ and $\nu\simeq0.839(9)$, $\beta/\nu\simeq0.152(2)$ for $F_C(\pi/2)$. 

For $R_b/a=3.4$, Fig. \ref{fig:app_z4_scaling_diagnostics} (c) and (d) use the same fixed period-4 component at $k=\pi/2$ and set $(\Delta/\Omega)_c=2.4158$. The collapse gives $\nu\simeq0.62(8)$, $\beta/\nu\simeq0.090(1)$ for $F_M(\pi/2)$ and $\nu\simeq0.70(10)$, $\beta/\nu\simeq0.094(5)$ for $F_C(\pi/2)$. For this cut, these exponent values are quoted only to compare the fixed-$\pi/2$ behavior of $F_M$ and $F_C$ under the same fitting protocol. In every inset, only $\nu$ and $\beta/\nu$ are optimized with $(\Delta/\Omega)_c$ fixed.
For both cuts, $F_C(\pi/2)$ grows with $\Delta/\Omega$ similar to $F_M(\pi/2)$, showing that the concurrence sum as well as magnetization contains information on the symmetry breaking related to period-4 modulation.

\section{Wave-vector comparison for $\mathbb{Z}_4$ transition}\label{app:z4_wavevector}
\begin{figure*}[t!]
\centering
\includegraphics[width=0.94\linewidth]{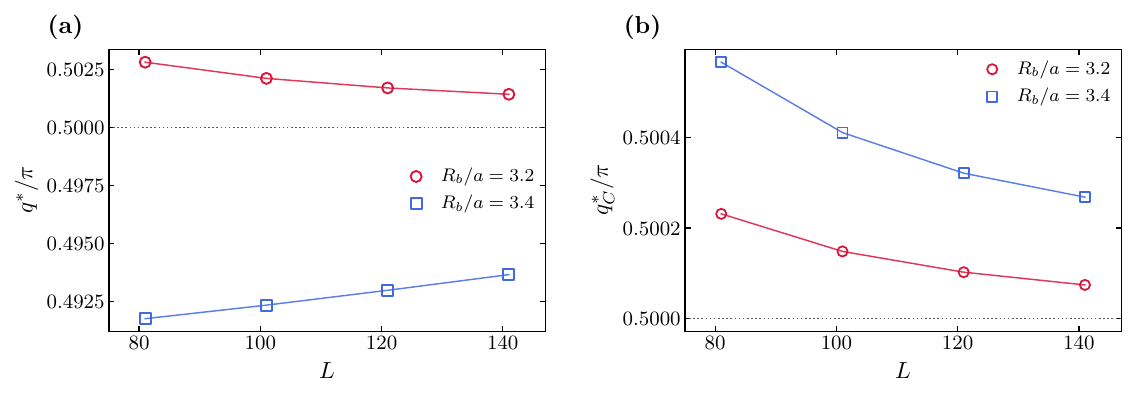}
\caption{Wave-vector comparison for the two $\mathbb{Z}_4$ transition cuts. In (a), $q^\ast/\pi$ is obtained from the density structure factor $S(k)$. In (b), $q_C^\ast/\pi$ is obtained from the Fourier amplitude $F_C(k)$ of the local concurrence sum. The horizontal dotted lines mark the period-4 value, i.e. $1/2$. The peak position of $S(k)$ separates the two cuts more clearly while the concurrence peak position remains very close to the period-4 value for both $\mathbb{Z}_4$ transition cuts.}
\label{fig:app_z4_concurrence_wavevector}
\end{figure*}

As a continuation of study on $\mathbb{Z}_4$ transitions, 
we here compare the peak position of the density structure factor $S(k)$ with that of the concurrence Fourier amplitude $F_C(k)$ as they look into different correlations, classical and quantum. Let $q^\ast$ be the wave vector at which $S(k)$ is maximal. We similarly define $q_C^\ast$ as the peak position of $F_C(k)$. Fig. \ref{fig:app_z4_concurrence_wavevector} compares these two peak positions in units of $\pi$ as a function of system size.

Fig. \ref{fig:app_z4_concurrence_wavevector} (a) shows that the peak position of the density structure $S(k)$ behaves differently in the two transition cuts. For $R_b/a=3.2$, the value of $q^\ast/\pi$ stays close to $1/2$, the period-4 value. On the other hand, for $R_b/a=3.4$, it remains slightly below $1/2$ and approaches $1/2$ slowly with increasing $L$. This is the finite-size trend behind the $q^\ast$ discussion in Sec. \ref{sec:z4_comparison}. In contrast, Fig. \ref{fig:app_z4_concurrence_wavevector} (b) shows that $q_C^\ast/\pi$ stays very close to $1/2$ for both cuts. Those peak positions therefore confirm that the concurrence structure is tied to the local period-4 modulation and that the distinction between the two cuts is more clearly resolved by $q^\ast$ of the density structure factor $S(k)$.

\begin{figure*}[t!]
  \includegraphics[width=1.0\linewidth]{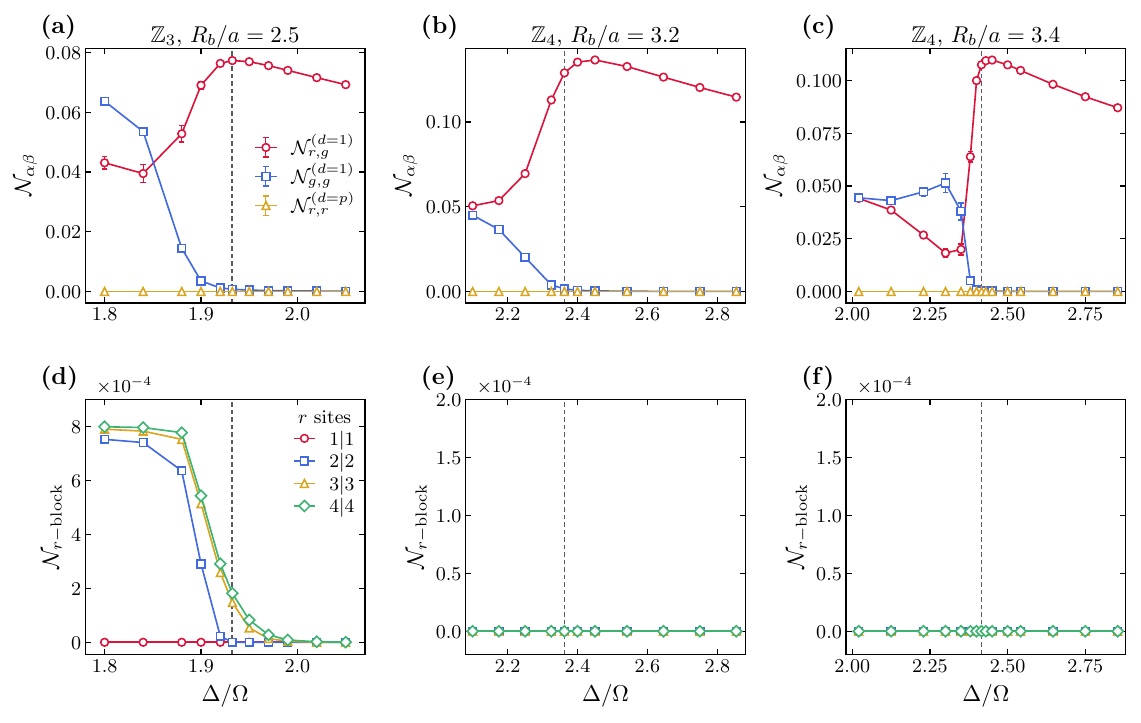}
  \caption{Representative negativity values (a) and (d): $\mathbb{Z}_3$ cut at $R_b/a=2.5$, (b) and (e): $\mathbb{Z}_4$ cut at $R_b/a=3.2$, (c) and (f): $\mathbb{Z}_4$ cut at $R_b/a=3.4$. The vertical dashed line marks the phase transition point for each case. (a)-(c) plots show the negativities evaluated for two selected sites as $\mathcal{N}_{r,g}^{d}$ Rydberg-ground,  $\mathcal{N}_{g,g}^{d}$ ground-ground, or  $\mathcal{N}_{r,r}^{d}$ Rydberg-Rydberg pairs of atoms at distance $d$. On the other hand, (d)-(f) plots show the negativities of a block of atoms using only the central Rydberg-state atoms. Each block has a total of $2M$ atoms in the bipartition $M:M$ ($M=1,2,3,4$). In (e) and (f), the block-negativity values are zero within numerical precision on the plotted scale.}
  \label{fig:app_negativity_scan}
\end{figure*}

\section{Negativity-based entanglement structure}\label{app:sublattice_negativity}
We here want to see whether a different measure of mixed-state entanglement shows a behavior similar to concurrence in entanglement structure. For this purpose, we evaluate the negativity \cite{vidal2002computable} of selected atoms in lattice using the definition introduced in Sec. \ref{sec:preliminaries}. We particularly choose atoms around the center of the Rydberg chain for the evaluation of negativities. 

In Fig. \ref{fig:app_negativity_scan} (a)-(c), $\mathcal{N}_{r,g}^{(d=1)}$ denotes the negativity for nearest-neighbor pairs of Rydberg-ground atoms, $\mathcal{N}_{g,g}^{(d=1)}$ for nearest-neighbor pairs of ground-ground atoms, and $\mathcal{N}_{r,r}^{(d=p)}$ for the pairs of Rydberg-Rydberg atoms separated by a distance of period $p$. The plotted values are averaged over three equivalent central pairs when available. Fig. \ref{fig:app_negativity_scan} (d)-(f) show block negativities evaluated for a total of $2M$ atoms all in the Rydberg-states with the bipartition $M:M$ ($M=1,2,3,4$).

The main trend in Fig. \ref{fig:app_negativity_scan} (a)-(c) is that the nearest-neighbor pair involving the atoms in the Rydberg-like state gives the largest two-site entanglement near and above the transition points. This is consistent with the behavior shown by concurrence profiles, for which the strongest local pair entanglement is attributed to pairs close to high-density sites. The nearest-neighbor pairs in both low density states and those in both high-density states but separated substantially by the distance of one period possess a much smaller negativity once the $\mathbb{Z}_4$ order is well established. 

Fig. \ref{fig:app_negativity_scan} (d)-(f) provide a separate comparison: they ask whether a block made only out of the high-density Rydberg sites carries appreciable bipartite negativity. This all-Rydberg-block-negativity is small in the plotted range for the $\mathbb{Z}_3$ cut and even gives zero within numerical precision for the two $\mathbb{Z}_4$ cuts. This again confirms that only a specific part of entanglement structure does not provide information well on the phase transition, as we showed with concurrence-based entanglement structure. Overall, the investigation on negativities reveals that the entanglement characteristics appear similar regardless of the specific entanglement measures.

\section{Entanglement structure of $\mathbb{Z}_2$ ordered phase}\label{app:z2_entanglement}
We here discuss  $\mathbb{Z}_2$-order phase more as $F_C(\pi)$ does not show a transition point near $\Delta_c/\Omega\simeq 1.08$ that is estimated from the density order parameter $F_M(\pi)$. In this phase, the next-nearest-neighbor concurrence appears significantly only after a larger $\Delta/\Omega\simeq 2.5$ than $\Delta_c/\Omega\simeq 1.08$. We here present more figures to explain the redistribution of local pairwise concurrence in the $\mathbb{Z}_2$ ordered phase.

\begin{figure}[h!]
    \includegraphics[width=1.0\linewidth]{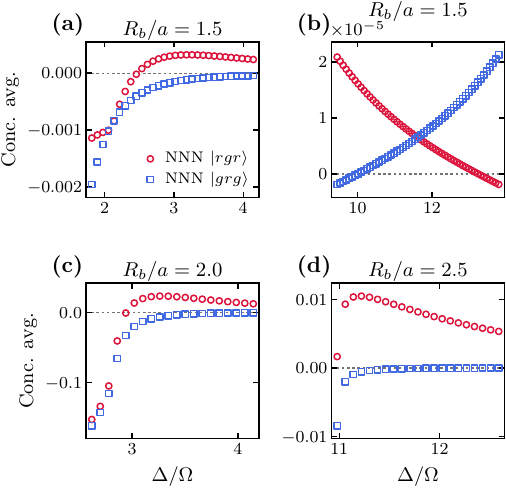}
    \caption{Plot of the next nearest neighbor (NNN) concurrence as a function of $\Delta/\Omega$ for various interaction strengths $R_b/a$. To examine its variation in detail, the quantity $\lambda_1-\lambda_2-\lambda_3-\lambda_4$ from Eq. (\ref{eq:concurrence}) is plotted. Here, NNN $|rgr\rangle$ and NNN $|grg\rangle$ denote the concurrence between two Rydberg-state atoms and two ground-state atoms, respectively, by excluding the middle atoms. The actual concurrence is obtained after applying the threshold in Eq. (\ref{eq:concurrence}).}
    \label{fig:NNN_concurrence}
\end{figure}

Fig. \ref{fig:NNN_concurrence} shows that the next nearest neighbor (NNN) concurrence between high-density (Rydberg) state atoms becomes appreciable only at larger detuning, above $\Delta/\Omega\simeq 2.5$. This explains why the $\mathbb{Z}_2$ profile in Fig. \ref{fig:chain_profile} can show a broken concurrence pattern at $\Delta/\Omega=4.25$, while $F_C(\pi)$ does not rise at the same detuning as $F_M(\pi)$.

\begin{figure*}[tbh!]
    \centering
    \vspace*{-24pt}
    \includegraphics[width=1.0\linewidth]{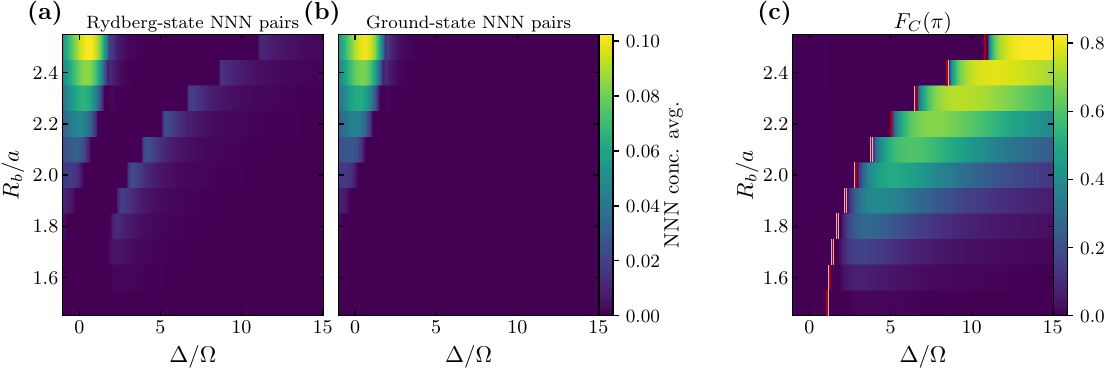} 
    \caption{Color maps of (a,b) local pairwise concurrence and (c) $F_C(\pi)$ for $N=121$. In (a), concurrence between next-nearest-neighbors in Rydberg state is plotted. In (b), concurrence between next-nearest-neighbors in ground state is plotted. In (c), $F_C(\pi)$ is plotted. The red guide in (c) indicates, for each $R_b/a$, the detuning where the magnetization order parameter $F_M(\pi)$ rises most rapidly.}
    \label{fig:concurrence_colormap}
    \vspace{3pt}
    \includegraphics[width=0.94\linewidth]{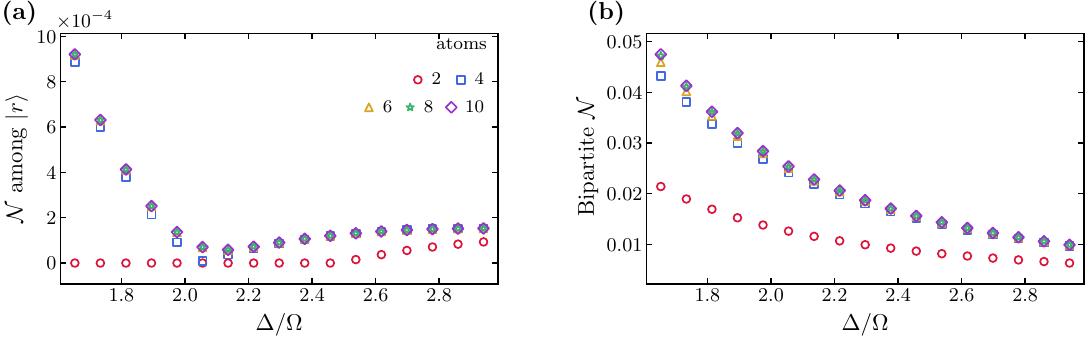}
    \caption{(a) Negativity of bipartitions formed from $2M$ high-density sites near the center of the chain ($M=1,2,3,4,5$) at $R_b/a=1.5$. (b) Bipartite negativity for adjacent blocks of $2M$ atoms over the same detuning window.}
    \label{fig:negativity}
\end{figure*}

The color maps in Fig. \ref{fig:concurrence_colormap} further show this behavior in the two-dimensional parameter space. The onset of the NNN concurrence and the corresponding growth of $F_C(\pi)$ occur on the larger-detuning side of the rapid-rise guide extracted from $F_M(\pi)$. Thus, there is an appreciable separation between the rapid rise of $F_M(\pi)$ and the substantial onset of the concurrence-based quantities in Fig. \ref{fig:concurrence_colormap}.
Finally, Fig. \ref{fig:negativity} shows that the feature visible in two-site concurrence becomes less sharp when larger blocks are considered. This may support a possible interpretation that the $\mathbb{Z}_2$ analysis here is a local pairwise-concurrence feature rather than a new phase transition.

\FloatBarrier
\bibliography{references}

\end{document}